\documentclass[conference]{IEEEtran}
\IEEEoverridecommandlockouts
\usepackage{cite}
\usepackage{amsmath,amssymb,amsfonts}
\usepackage{algorithmic}
\usepackage{graphicx}
\usepackage{subfigure}
\usepackage{array}
\usepackage{tabularx}
\usepackage{textcomp}
\usepackage{xcolor}
\usepackage{url}
\def\BibTeX{{\rm B\kern-.05em{\sc i\kern-.025em b}\kern-.08em
    T\kern-.1667em\lower.7ex\hbox{E}\kern-.125emX}}

\begin{document}

\title{Enhanced Optimal Power Flow Based Droop Control in MMC-MTDC Systems\\
\thanks{This work is supported by CRESYM project Harmony (\protect\url{https://cresym.eu/harmony/}).}}

\author{\IEEEauthorblockN{Hongjin Du, Rashmi Prasad, Aleksandra Lekić, Pedro P. Vergara, Peter Palensky}
\IEEEauthorblockA{\textit{Department of Electrical Sustainable Energy},
\textit{Delft University of Technology} \\
Delft, Netherlands \\
\{h.du, r.prasad, a.lekic, p.p.vergarabarrios, p.palensky\}@tudelft.nl}}

\maketitle

\begin{abstract}
Optimizing operational set points for modular multilevel converters (MMCs) in Multi-Terminal Direct Current (MTDC) transmission systems is crucial for ensuring efficient power distribution and control. This paper presents an enhanced Optimal Power Flow (OPF) model for MMC-MTDC systems, integrating a novel adaptive voltage droop control strategy. The strategy aims to minimize generation costs and DC voltage deviations while ensuring the stable operation of the MTDC grid by dynamically adjusting the system operation points. The modified Nordic 32 test system with an embedded 4-terminal DC grid is modeled in Julia and the proposed control strategy is applied to the power model. The results demonstrate the feasibility and effectiveness of the proposed droop control strategy, affirming its potential value in enhancing the performance and reliability of hybrid AC-DC power systems.
\end{abstract}


\section{Introduction}
The growing adoption of variable renewable energy sources (VRES) introduces challenges such as variability and unpredictability which can destabilize traditional power grids. High Voltage Direct Current (HVDC) technology has emerged as a promising solution as it enables efficient long-distance transmission and integration of VRES. The implementation of the long-distance HVDC connections has been successful, with numerous examples already established worldwide, as documented in \cite{Kim2009HVDCTP}. To integrate remotely located energy sources with the AC grid, MTDC grids have been proposed as an alternative to conventional HVDC systems. At the same time, MMCs have emerged as the most popular Voltage Source Converters (VSCs) due to their controllability and modularity. The MMC-MTDC system has been widely researched due to its significant features, such as the ability to include independent control of active and reactive power and interconnection of weak AC systems or even passive networks \cite{867439}. 

Generally, research on control in MTDC systems can be categorized into master-slave control and voltage droop control \cite{6588621}. Master-slave control provides accurate power-sharing control but requires fast communication to send the active power set-point to the power-controlled converters periodically for any changes in generated power. However, it exhibits low reliability stemming from the potential for single-point failure of the master converter and challenges in expansion. Conversely, voltage droop control uses local voltage and current measurements to control the converter's output, and the droop coefficient determines the power-sharing among the converters without a central controller or communication. However, inaccurate power sharing can result in overload or underutilization of some converters, potentially limiting the economic viability and scalability of power systems. To address this, several studies have proposed adaptive and nonlinear droop control methods that can achieve accurate current sharing and improved voltage regulation, even in the presence of practical factors such as sensor calibration errors and cable resistances\cite{7967860, 8616804, 9594778, 7484690}.

In addition to dynamic responses, focusing on a local control strategy from a power flow perspective is crucial for optimizing operational efficiency and system stability. In \cite{Beerten2011VSCMS}, a distributed DC voltage control approach is introduced, where multiple converters can collaboratively regulate the DC system voltage. After a fault event, the voltage droop-controlled converters can adjust to new operating points influencing the entire system. In \cite{stojkovic2020adaptive}, the impact of converter droop settings and MTDC network topology on power sharing in MTDC grids is studied, presenting an analytical tool for evaluating the effect of droop control settings on steady-state voltage deviations and power-sharing after a converter outage. In \cite{lekic2020initialisation}, the power flow solution serves as the initialization for a hybrid AC-DC power system, providing a starting point for electromagnetic or harmonic stability analyses. 

The current research on MMC stations within the OPF context often simplifies the AC-DC interconnection aspect and neglects system stability evaluation. This oversight suggests that existing models may not fully capture the complexities of AC-DC interconnection. Many existing studies employ fixed droop coefficient control schemes and prioritize minimizing fuel costs in OPF calculations. However, this approach presents several limitations. Fixed droop coefficients cannot adapt to system disturbances or effectively respond to voltage fluctuations. Moreover, frequent adjustments to droop coefficients may destabilize the system due to over-adjustment and control interactions \cite{9178970}. While reducing power loss and operation costs are important objectives for interconnected AC-DC grids, focusing solely on minimizing fuel costs can lead to larger voltage variations. In reality, voltage stability is vital for ensuring the stability of a large-scale power system. Therefore, it is crucial to achieve a balance between system cost and voltage stability when designing control strategies for AC-DC interconnected grids.

Motivated by the aforementioned problems, this paper proposes an enhanced OPF-based voltage droop control strategy for MMC-MTDC systems. The remainder of this paper is structured as follows. The configuration of MMC stations and previous voltage droop control methods are reviewed in Section \uppercase\expandafter{\romannumeral2}. The advanced OPF-based voltage droop control strategy is presented in Section \uppercase\expandafter{\romannumeral3}. The feasibility of the proposed droop control scheme and the stability of the test system is verified by a case study in Section \uppercase\expandafter{\romannumeral4}, and Section \uppercase\expandafter{\romannumeral5} draws conclusions.
\section{Configuration of MMC Stations and voltage droop control}
\begin{figure}[htbp]
\centerline{\includegraphics[scale=0.8]{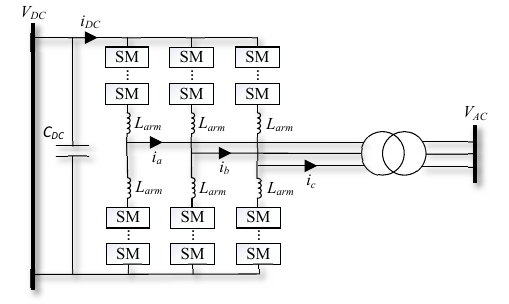}}
\caption{Overview of investigated configuration of MMC.}
\label{mmc}
\end{figure}

MMC stations serve as vital nodes in MTDC systems, facilitating the conversion of AC to DC and vice versa. The fundamental structure of an MMC comprises multiple half-bridge sub-modules connected in series to form arms and then further combined to create converter cells. These cells operate with pulse-width modulation (PWM) techniques, enabling precise control over voltage and current waveforms. The configuration of MMC stations is shown in Fig.~\ref{mmc}. By using multiple levels of control to approximate a sine wave, MMCs can produce smoother output waveform compared to conventional VSCs. The inherent characteristics of MMCs can help mitigate harmonics without the need for extra filtering, thus potentially reducing system complexity and cost. 

A typical control structure of a MMC station involves a double-loop control system. The outer loop, also known as the secondary control, deals with the DC voltage and the active power according to the droop characteristic. The active power injection can be represented by 
\begin{equation}
    P_{dc,i}-P_{dc,0}+1/k_{droop}\left(U_{dc, i}-U_{dc,0}\right)=0,
    \label{droopfunction}
\end{equation}
where $k_{droop} = \Delta U_{dc} / \Delta P_{dc}$; $P_{dc, i}$ and $P_{dc,0}$ are the actual and reference DC power for droop control; and $U_{dc, i}$ and $U_{dc,0}$ are the actual and reference voltage, respectively. In an ideal lossless grid, the distribution of imbalanced power is allocated among droop-controlled converters proportionally to their respective droop gains. Consequently, the operator can adjust the droop gain to ensure appropriate power-sharing following a disturbance. The parameter $k_{droop}$ directly influences both the power distribution and the DC control performance of the entire system. Given that frequent adjustments can impact system stability, selecting an appropriate value beforehand and subsequently adjusting the reference value is crucial.

\section{Proposed Enhanced OPF-based Voltage Droop Control}
In this section, an enhanced voltage droop control strategy based on power flow algorithms is proposed to address the converter control influence on the steady-state model. The format of the AC/DC power flow models is outlined in \cite{5589968}. Consequently, the proposed optimization problem is formulated based on this.

\subsection{Formulation of the Optimization Problem}
The optimization problem is a two-stage problem minimizing both generation cost and DC voltage deviation. The objectives are defined as
\begin{align}
Obj_1 = \min \sum_{i=1}^M (\alpha_iP_{Gi}^2 + \beta_iP_{Gi} + \gamma_i),\\
Obj_2 = \min \sum_{i=1}^N (U_{dc,i} - U_{dc,N})^2 ,
\end{align}
where $Obj_1$ and $Obj_2$ minimize generation cost and DC voltage deviation, respectively. $P_{Gi}$ represents the active power generation at bus $i$. The coefficients $\alpha_i$, $\beta_i$, and $\gamma_i$ are the cost parameters of generators. $U_{dc, i}$ and $U_{dc, N}$ are the desired voltage and rated voltage at DC node $i$. It should be noted that the problem is not set as a multi-objective problem because voltage stability has a higher priority. 

For AC network, the power flow equations can be written as
\begin{align}
P_{i} = U_i \sum\limits_{j=1}^n U_j[G_{ij}cos(\delta_i-\delta_j)+B_{ij}sin(\delta_i-\delta_j)],\\
Q_{i} = U_i \sum\limits_{j=1}^n U_j[G_{ij}sin(\delta_i-\delta_j)-B_{ij}cos(\delta_i-\delta_j)],\\
P_{Gi} - P_{Di} - P_{i} = 0,\quad
Q_{Gi} - Q_{Di} - Q_{i} = 0,
\end{align}
where $P_{i}$, $P_{Di}$, and $Q_{i}$, $Q_{Di}$ are the active and reactive power of the node and demand, respectively. Additionally, $Q_{Gi}$ is the reactive power generation and $U_i$ indicates the voltage magnitude. $G_{ij}$ and $B_{ij}$ denote the conductance and susceptance between bus $i$ and $j$. $\delta_i$ and $\delta_j$ represent the voltage angles.

The DC grid model is defined by the equations: 
\begin{align}
I_{dc,i}=\sum\limits_{j=1,j \neq i}^n Y_{dc,ij} (U_{dc,i}-U_{dc,j}),\\
P_{dc,i} = 2U_{dc,i}I_{dc,i},
\end{align}
where $Y_{dc,ij}$ is the admittance between buses $i$ and $j$ in the DC grid; $P_{dc,i}$ and $I_{dc,i}$ are the active power and current injected in the DC network at DC node $i$. 

The converter loss, denoted as $P_{loss}$, consists of three components: no-load losses, linear losses, and quadratical losses of the converter current $I_{c,i}$. The per unit coefficients $a$, $b$, and $c$ are derived from empirical data as shown in Table~\ref{powerloss}. 
\begin{align}
P_{loss,i} = a+b*I_{c,i}+c*I_{c,i}^2\\
P_{c,i}^2 + Q_{c,i}^2 - 3U_{c,i}I_{c,i} = 0
\end{align}

\begingroup
\begin{table}[t!] 
\caption{Power Loss Coefficients} \label{powerloss}
\centering
\footnotesize 
\renewcommand{\arraystretch}{1.25}
\begin{tabularx}{0.45\textwidth}{ >{\raggedright\arraybackslash}X >{\centering\arraybackslash}X >{\centering\arraybackslash}X >{\centering\arraybackslash}X }
 
\textbf{Coefficients}&\textbf{a}&\textbf{b}&\textbf{c}\\ [1pt]
 
\hline
Rectifier & 0.011 & 0.003 & 0.004\\[1pt]
Inverter & 0.011 & 0.003 & 0.007\\[1pt]
\hline
\end{tabularx}
\end{table}
\endgroup

The limits on the voltage, generation power, and DC power are reinforced by the constraints:
\begin{align}
     U_i^{min} \leq U_i \leq U_i^{max}, \quad 
    \delta_i^{min} \leq \delta_i \leq \delta_i^{max}, \\\quad
     P_{Gi}^{min} \leq P_{Gi} \leq P_{Gi}^{max}, 
     Q_{Gi}^{min} \leq Q_{Gi} \leq Q_{Gi}^{max},\\ \quad
     P_{dc,i}^{min} \leq P_{dc,i} \leq P_{dc,i}^{max}.
\end{align}

\subsection{Proposed Droop Control Strategy}
In this paper, the active power of DC nodes is adjusted based on the voltage deviation according to (\ref{droopfunction}). For DC nodes in droop-control mode, the droop coefficient $k_{droop}$ is positive, while for active power-controlled nodes, $k_{droop}$ equals zero.

\begin{figure}[htbp]
\centering
\includegraphics[scale=0.8]{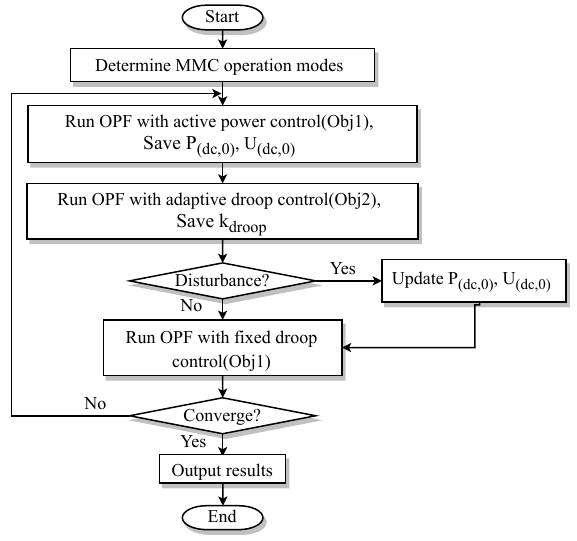}
\caption{Flowchart of the proposed strategy.}
\label{flowchart}
\end{figure}

The flowchart of the proposed strategy is illustrated in Fig.~\ref{flowchart}. Initially, the system configures the MMC operation modes for active power control, with the objective function $Obj_1$ aimed at minimizing generation costs.

Gaining the active power and voltage set points at DC buses, we implement the VSC control mode as adaptive droop control, where the droop coefficient $k_{droop}$ is treated as a variable. Subsequently, we execute a second round of optimization aimed at minimizing the voltage deviation at each DC bus. It's essential to emphasize that these two stages of optimization cannot be combined due to the alterations in constraints and the prioritization of voltage stability requirements.

As mentioned in \cite{9178970}, adapting the droop coefficients may influence the stability of the AC-DC system, as it may lead to control interactions, where one terminal's adjustment affects the voltage at another terminal, potentially causing oscillations or instability if not properly coordinated. To avoid such negative impacts on the system, the droop coefficients are set as fixed at the beginning, whereas the reference values are updated after the disturbance. In the end, the final round of optimization is conducted with the objective function $Obj_1$, and the droop equation is updated as follows.
\begin{equation}
    P_{dc,i}-P_{dc,new}+1/k_{droop}\left(U_{dc, i}-U_{dc,new}\right)=0 
\end{equation}
      
\section{Case study}
\begin{figure}[htbp]
    \centering
    \subfigure[Nordic 32 test system]{
        \includegraphics[width=0.4\textwidth]{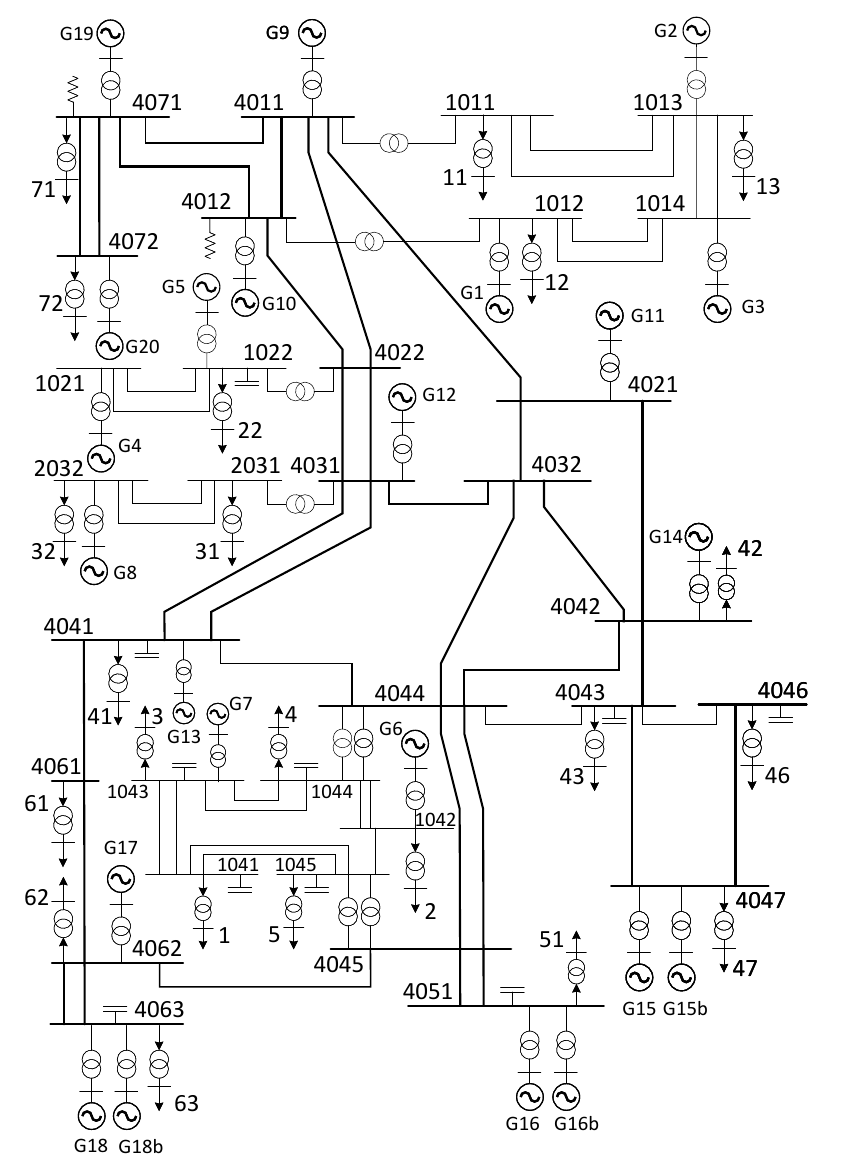}}
    \subfigure[MTDC system]{
        \includegraphics[width=0.25\textwidth]{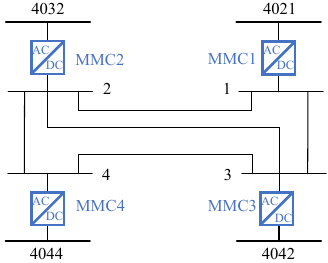}}
    \caption{Nordic test system with a 4-terminal MTDC system \cite{8950069, van2015test}.}
    \label{nordic}
    \vspace*{-5mm}
\end{figure}

The modified Nordic 32 test system incorporating an embedded 4-terminal DC grid which is shown in Fig.~\ref{nordic}, is modeled for OPF simulation using Matpower format. The case system parameters utilized in this paper are sourced from \cite{8950069} and \cite{van2015test}. The nominal power of the whole system is $S_{nom} = 100$ MVA while the DC nominal voltage is $V_{nom} = 200$ kV. 

Following that, the proposed control strategy is executed, building upon the power model as its cornerstone. The optimization process follows the format of the Julia package PowerModelsACDC.jl \cite{8636236} to determine the operating point. The Ipopt solver, embedded in the JuMP toolbox, is then utilized to solve the nonlinear optimization problem. Subsequently, the obtained results are cross-validated in EMTP to assess the system's stability under the proposed control strategy.

Three scenarios have been chosen as test cases for the system based on empirical data:
\begin{enumerate}
    \item \textbf{Scenario 1:} The system operates under normal condition.
    \item \textbf{Scenario 2:} A fault arises in generator 16, resulting in its outage.
    \item \textbf{Scenario 3:} A fault occurs in MMC 4, leading to its disconnection.
\end{enumerate}
Scenarios 2 and 3 present opportunities to evaluate the effectiveness of the droop control mode and assess the performance of the power flow model under unbalanced operating conditions. Three different control modes are tested for each scenario, which include:
\begin{enumerate}
\item Active power control,
\item Adaptive voltage droop control (with $k_{droop}$ set as a variable and a fixed reference value), 
\item Proposed voltage droop control.
\end{enumerate}

Solving the OPF problem across various scenarios yields the droop coefficients for both the proposed strategy and adaptive droop control, as illustrated in Table \ref{kdroop} and Table \ref{kdroop adapt}, respectively. To maintain system stability, the range for the droop coefficient ($k_{droop}$) is carefully set between 0.001 and 0.5. This range ensures a balanced trade-off between dynamic response and voltage regulation. 

In Scenario 2, following the contingency event, the system demonstrates resilience as the proposed droop control strategy successfully finds the optimal solution using pre-calculated $k_{droop}$ values, while adjusting merely $P_{ref}$ and $U_{ref}$. However, Scenario 3 presents a substantial challenge following a converter outage. As the limitations of the converter control capability and the voltage stability requirements, the power disturbance is too significant for droop control adjustments alone. Consequently, the feasibility of OPF calculations becomes compromised. Following the process outlined in Fig.~\ref{flowchart}, the control strategy reverts to its initial state, initiating a new iteration of droop coefficient calculation. The divergent behaviors observed in Scenario 2 and Scenario 3 highlight the sensitivity and robustness of the control strategy to different system conditions and fault scenarios.

\begingroup
\begin{table}[t!] 
\caption{$k_{droop}$ in different scenarios under proposed droop control} \label{kdroop}
\centering
\footnotesize 
\renewcommand{\arraystretch}{1.25}
\begin{tabularx}{0.45\textwidth}{ >{\raggedright\arraybackslash}X >{\centering\arraybackslash}X >{\centering\arraybackslash}X >{\centering\arraybackslash}X }
 
\textbf{$k_{droop}$} & \textbf{Scenario 1} & \textbf{Scenario 2} & \textbf{Scenario 3}\\ [1pt]
 
\hline
MMC 1 & 0.4999 & 0.4999 & 0.5000\\[1pt]
MMC 2 & 0.0010 & 0.0010 & 0.5000\\[1pt]
MMC 3 & 0.3003 & 0.3003 & 0.3409\\[1pt]
MMC 4 & 0.4967 & 0.4967 & -\\[1pt]
\hline
\end{tabularx}
\end{table}
\endgroup

\begingroup
\begin{table}[t!] 
\caption{$k_{droop}$ in different scenarios under adaptive droop control} \label{kdroop adapt}
\centering
\footnotesize 
\renewcommand{\arraystretch}{1.25}
\begin{tabularx}{0.45\textwidth}{ >{\raggedright\arraybackslash}X >{\centering\arraybackslash}X >{\centering\arraybackslash}X >{\centering\arraybackslash}X }
 
\textbf{$k_{droop}$} & \textbf{Scenario 1} & \textbf{Scenario 2} & \textbf{Scenario 3}\\ [1pt]
 
\hline
MMC 1 & 0.0979 & 0.0678 & 0.0676 \\[1pt]
MMC 2 & 0.0775 & 0.0454 & 0.0631 \\[1pt]
MMC 3 & 0.5000 & 0.0010 & 0.0010\\[1pt]
MMC 4 & 0.0010 & 0.0146 & -\\[1pt]
\hline
\end{tabularx}
\end{table}
\endgroup

 The final optimization results of minimizing generation cost are shown in Table~\ref{objective}. Notably, active power control demonstrates better cost-effectiveness compared with droop control due to its unconstrained nature in power regulation. The absence of constraints enables active power control to achieve lower generation costs, showing its efficiency in optimizing economic performance.
 The gaps between adaptive droop control and proposed droop control are marginal, typically lower than 0.4\%, which demonstrates the reliability of the proposed strategy in achieving comparable outcomes. Such consistency across control methods not only reaffirms the viability of the proposed strategy but also highlights its potential for seamless integration into existing systems. 
\begingroup
\begin{table}[t!] 
\caption{Objective values in different scenarios} \label{objective}
\centering
\footnotesize 
\renewcommand{\arraystretch}{1.25}
\begin{tabularx}{0.48\textwidth}{ >{\raggedright\arraybackslash}p{3cm}  >{\centering\arraybackslash}X >{\centering\arraybackslash}X >{\centering\arraybackslash}X }
 
{\textbf{Objectives ($\times 10^{6}$ MW)}} & \textbf{Scenario 1} & \textbf{Scenario 2} & \textbf{Scenario 3}\\ [1pt]
 
\hline
Active Power Control & 8.78 & 9.30 & 8.78 \\[1pt]
Adaptive Voltage Droop & 8.90 & 9.45 & 8.90 \\[1pt]
Proposed Voltage Droop & 8.94 & 9.50 & 8.90\\[1pt]
\hline
\end{tabularx}
\end{table}
\endgroup

\begin{figure}[htbp]
    \centering
    \subfigure[Scenario 1: Normal condition]{
    \includegraphics[width=0.35\textwidth]{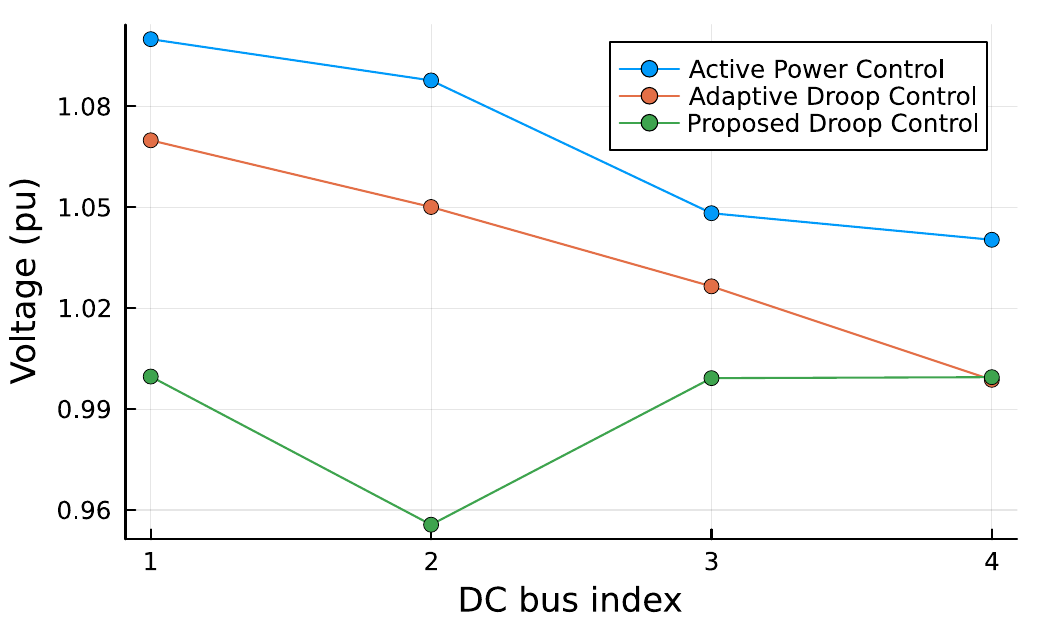}}
    \subfigure[Scenario 2: Generator 16 outage]{
    \includegraphics[width=0.35\textwidth]{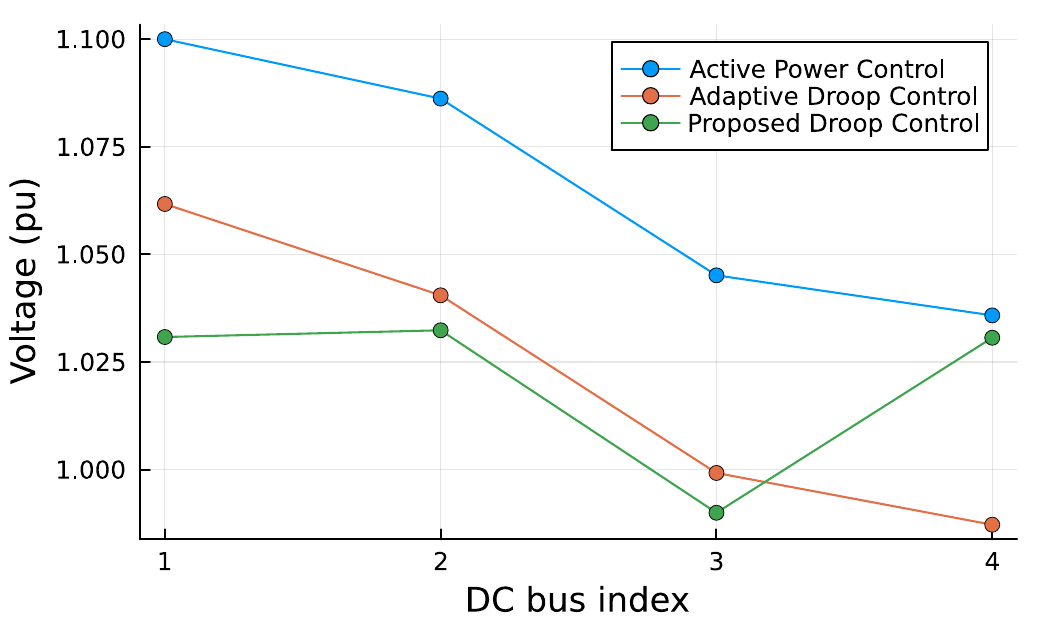}}
    \subfigure[Scenario 3: MMC 4 disconnection]{
    \includegraphics[width=0.35\textwidth]{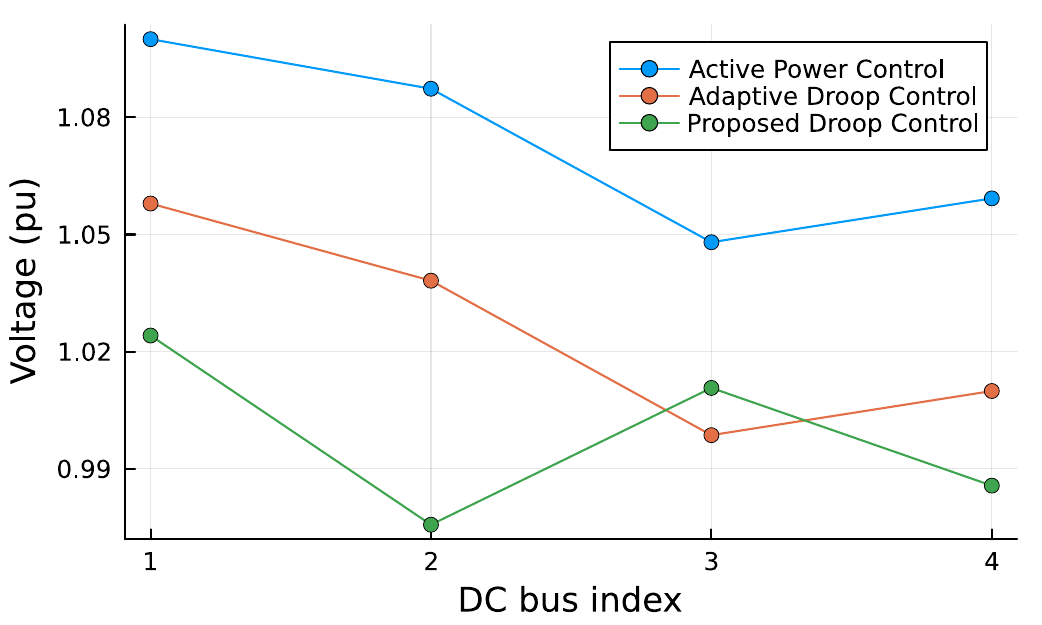}}
    \caption{DC voltage in different scenarios.}
    \label{fig:voltage}
\end{figure}

In Fig.~\ref{fig:voltage}, voltage set points at different DC buses are depicted under various scenarios. Subfigure (a) illustrates the DC voltage under normal operating conditions, while subfigures (b) and (c) represent scenarios involving generator outage and MMC outage, respectively. Comparing the voltage regulation performance across scenarios, it becomes apparent that the proposed control strategy offers superior voltage regulation capabilities, primarily attributable to the implementation of Objective 2. As shown in Table~\ref{voltage deviation}, where the voltage deviation is calculated by $\sum\limits_{i=1}^N (U_{dc,i} - U_{dc, N})^2$, the DC voltage tends to converge closer to the desired 1 pu reference under the proposed control strategy, indicating a more resilient power system that is better equipped to withstand varying operating conditions and disturbances. In contrast, both the adaptive droop control and active power control strategies exhibit voltage levels that are consistently closer to the operational limits.    

\begingroup
\begin{table}[t!] 
\caption{Voltage deviation in different scenarios} \label{voltage deviation}
\centering
\footnotesize 
\renewcommand{\arraystretch}{1.25}
\begin{tabularx}{0.48\textwidth}{ >{\raggedright\arraybackslash}p{3cm}  >{\centering\arraybackslash}X >{\centering\arraybackslash}X >{\centering\arraybackslash}X }
 
{\textbf{Voltage deviation(pu)}} & \textbf{Scenario 1} & \textbf{Scenario 2} & \textbf{Scenario 3}\\ [1pt]
 
\hline
Active Power Control   & 0.0217 & 0.0207 & 0.0234 \\[1pt]
Adaptive Voltage Droop & 0.0081 & 0.0056 & 0.0049 \\[1pt]
Proposed Voltage Droop & 0.0020 & 0.0030 & 0.0015\\[1pt]
\hline
\end{tabularx}
\end{table}
\endgroup

\section{Conclusion}
This paper proposes an enhanced OPF-based droop control strategy tailored for MMC-MTDC systems. The primary objective is to minimize generation costs and voltage deviations by optimizing system operation points. Additionally, the strategy aims to maintain voltage levels at DC buses close to nominal values, thereby allowing converters with larger power margins to absorb a greater share of power mismatches and effectively mitigate contingencies.

Taking the Nordic 32 system integrated with a four-terminal MTDC grid as a test case, the optimization problem is implemented in PowerModelsACDC. Through a series of scenarios, this paper demonstrates the effectiveness of the proposed control strategy, showcasing its practical implications to enhance system performance and reliability. A possible extension for future work is to incorporate uncertainties from variable renewable energy sources and loads in order to design more robust control strategies.

\bibliographystyle{IEEEtran}
\bibliography{manual.bib}

\end{document}